\documentstyle[aps,twocolumn,prb,epsfig]{revtex}

\begin{document}

\title{Pairing, Charge, and Spin Correlations in the Three-Band
Hubbard Model} 
\author{Z. B. Huang and H. Q. Lin
   \thanks{Corresponding author, email address: hqlin@phy.cuhk.edu.hk}}
\address{ Department of Physics, Chinese University of Hong Kong, Hong Kong,
China }
\author {J. \ E. \ Gubernatis}
\address{ Theoretical Division, Los Alamos National Laboratory,
	Los Alamos, NM 87545 }

\date{\today}

\maketitle
\vskip -0.5truecm

\begin{abstract}
Using the Constrained Path Monte Carlo (CPMC) method, we simulated the
two-dimensional, three-band Hubbard model to study pairing, charge, and spin
correlations as a function of electron and hole doping and the Coulomb
repulsion $V_{pd}$ between charges on neighboring Cu and O lattice
sites. As a function of distance, both the $d_{x^2 - y^2}$-wave and
extended s-wave pairing correlations decayed quickly. In the
charge-transfer regime, increasing $V_{pd}$ decreased the long-range
part of the correlation functions in both channels, while in the
mixed-valent regime, it increased the long-range part of the s-wave
behavior but decreased that of the d-wave behavior. Still
the d-wave behavior dominated.  At a given doping, increasing $V_{pd}$
increased the spin-spin correlations in the charge-transfer regime but
decreased them in the mixed-valent regime. Also increasing $V_{pd}$
suppressed the charge-charge correlations between neighboring Cu and O
sites. Electron and hole doping away from half-filling was accompanied
by a rapid suppression of anti-ferromagnetic correlations.
\end{abstract}

\narrowtext

\section{Introduction}

In this paper, we will report the results of a quantum Monte Carlo
(QMC) study of the ground-state properties of the two-dimensional,
three-band Hubbard model. Of the three simple electronic models
commonly studied as possible models of the cuprate superconducting
materials, the three-band Hubbard model has been the least intensively
studied, partially because of the general belief that its low energy
excitation spectrum is similar to the other two. Indeed in the strong
coupling limit, both the one-band and three-band Hubbard models have
the t-J model as an approximate limit. However, there still remains
some controversy about whether one-band models, like the Hubbard and
t-J models, are adequate to describe the low energy physical
properties of the cuprate superconductors. One of our objectives was
to study the possible existence of superconductivity in the three-band
model in regions where model parameters are physical as opposed to
regions where asymptotic models are clearly more appropriate. A focus
of our ground state study is the effect of the inclusion of $V_{pd}$,
a repulsive Coulomb interaction between charges on neighboring Cu and
O lattice sites.

The most solid information about possible superconductivity in the
one-band Hubbard model has come from a series of QMC calculations. For
instance, using a finite temperature QMC method, White et
al. \cite{white1} found an attractive effective pairing interaction in
the $d_{x^2 - y^2}$ and extended s-wave channels. Moreo and Scalapino
\cite{moreo} subsequently found pairing correlations but also found
that they did not increase when the lattice size was increased. This
result, suggesting the absence of off-diagonal long range order, was
consistent with an earlier QMC study by Imada and
Hatsugai.\cite{imada}

The fermion sign problem limits almost all QMC calculations of the
Hubbard model to small system sizes and finite temperature simulations
to high temperatures.  These limitations had left open the possibility
that superconductivity still lurked at larger systems sizes and lower
temperatures. Recently, a new zero temperature QMC method, the
Constrained Path Monte Carlo (CPMC) method,\cite{zhang1} was
developed to get around the sign problem. Using this method, Zhang et
al. \cite{zhang2} calculated ground-state pairing correlation
functions for the Hubbard model as a function of the distance and
found that as the system size or the interaction strength is increased
the magnitude of the long-range part of the pairing correlation
functions vanished for both the $d_{x^2 - y^2}$ and extended s-wave
channels. Although this method produces an approximate solution, its
results, together with the null results from previous QMC studies, are
very discouraging for finding superconductivity in the one-band
Hubbard model.

At finite temperature, past numerical work on the three-band model has
included several sets of QMC simulations,
\cite{scalettar1,scalettar2,dopf1,dopf2,dopf3,dopf4}
focusing on the magnetic
superconducting and insulating properties of the model.  As for the
QMC simulations of the Hubbard model, the sign problem limited these
studies to relatively high temperatures and small systems. In fact the
sign problem for the three-band model is probably more severe
than for the one band model. At least the one band model lacks a sign
problem at half filling. In general, an anti-ferromagnetic state is
found at half-filling which is strongly suppressed upon
doping. Attractive interactions between pairs were found with a
spectrum of results: Dominance of extended s-wave and d-wave pairing
have been separately reported, leading to claims of extended s-wave
and d-wave ODLRO. These results remain controversial.

At zero temperature, past numerical work on the three-band model has
mainly consisted of exact diagonalization
computations,\cite{ogata}$^{- }$\cite{stephan} where calculations of
hole binding energies was emphasized, and several QMC computations
\cite{kuroki,frick,comments} where pairing calculations were emphasized. 
The exact diagonalization studies
unequivocally established that holes can bind. The QMC studies,
established the existence of an extended s-wave and $d_{x^2-y^2}$ wave
attractive pairing interaction, with one claim of of no evidence of
s-wave superconductivity. More recently, a CPMC study by Guerrero and
Gubernatis \cite{guerr} confirmed the exact diagonalization result that
holes bind but found that increasing system size tended to decrease the long
range part of the pairing correlations. In contrast to the exact
diagonalization work, this QMC study found hole binding in the
absence of a Coulomb repulsion $V_{pd}$ between charge on neighboring
Cu and O sites. The exact diagonalization studies found that hole
binding required an unphysically large value of $V_{pd}$.

In the work reported here, we applied the CPMC method to the
three-band Hubbard model and computed the static pairing, charge, and
spin correlation functions for systems with $6 \times 6$ unit
cells. We set the hopping and on-site Coulomb parameters as expected
physically relevant values and studied the properties of the systems
as a function of charge-transfer energy, electron and hole doping, and
$V_{pd}$ through a range of physically relevant values. The
consequences of varying these parameters for the most part depended on
whether the value of the charge-transfer energy placed the model in a
charge-transfer or mixed valent regime. The cuprate superconductors
are believed to be in the charge-transfer regime.

The remainder of our report is organized as follows: In Section
\ref{HAMILTON}, we define the Hamiltonian and the physical quantities
calculated and discuss the choice of model parameters. In Section
\ref{CPMC} we briefly describe the CPMC method, and then in Section
\ref{RESULTS}, we present our numerical results.  Finally in Section
\ref{CONCLUSIONS}, we discuss in detail our main conclusions.

\section{\label{HAMILTON} Three-band Hubbard Hamiltonian }

As proposed by Emery,\cite{emery} the three-band model mimics the
CuO$_2$ layer in the cuprate superconductors by having one Cu and two
O atoms per unit cell, with the Cu atoms arranged on a square lattice
and the O atoms centered on the edges of the square unit cells. In
this layer, Emery assumed that the relevant orbitals are just those of
copper $\ 3d_{x^2-y^2}$ and oxygen $2p_x $ and $2p_y$.

The Hamiltonian has the form:
\begin{eqnarray}
H &=& \sum_{<j,k> \sigma} t_{pp}^{jk} (p^{\dagger}_{j \sigma}p_{k\sigma}^{}
 +p^{\dagger}_{k\sigma}p_{j \sigma}^{} )
 +\epsilon_p \sum_{j\sigma} n^{p}_{j\sigma}
 +U_p\sum_j n_{j\uparrow}^p n_{j\downarrow}^p  \nonumber   \\
& & \hspace{2cm} + \epsilon_d \sum_{i\sigma} n^{d}_{i\sigma}
 + U_d\sum_i n_{i\uparrow}^d n_{i\downarrow}^d   \\
& & \hspace{1cm} + V_{pd}\sum_{<i,j>} n_{i}^d n_{j}^p \nonumber
 + \sum_{<i,j>\sigma} t_{pd}^{ij} (d^{\dagger}_{i\sigma}p_{j\sigma}^{}
 + p^{\dagger}_{j\sigma}d_{i\sigma}^{})   \nonumber
\end{eqnarray}
In writing the Hamiltonian, we adopted the convention that the
operator $d^\dagger_{i\sigma}$ creates a {\it hole} with spin $\sigma$
at a Cu $3d_{x^2 - y^2}$ orbital and $p^\dagger_{j\sigma}$ creates a
{\it hole} with spin $\sigma$ in an O $2p_x$ or $2p_y$ orbital.  $U_d$
and $U_p$ are the Coulomb repulsions at the Cu and O sites,
$\epsilon_d$ and $\epsilon_p$ are the corresponding orbital energies,
and $V_{pd}$ is the nearest neighbor Coulomb repulsion. As written,
the model has a Cu-O hybridization $t_{pd}^{ij} = \pm t_{pd}$ with the
minus sign occurring for $j = i + \hat{x}/2 $ and $j = i - \hat{y}/2 $
and also hybridizaton $ t_{pp}^{jk} = \pm t_{pp} $ between oxygen
sites with the minus sign occurring for $k = j - \hat{x}/2 -
\hat{y}/2$ and $k = j + \hat{x}/2 + \hat{y}/2$. These phase
conventions are illustrated in Fig.~1.

The values of the parameters in the Hamiltonian have been estimated by
a number of different constrained density functional and quantum
cluster calculations.\cite{hybertsen}${^-}$\cite{martin} In
electron-volts reasonable ranges for these values seem to be: $t_{pd}=
1.3-1.6$, $U_d=8.5-10.5$, $\epsilon=\epsilon_p-\epsilon_d=3.6$,
$U_p=4.0-7.5$, $t_{pp}=0.65$, and $V_{pd}=0.6-1.2$. Taken together, these
estimates define a reasonably limited range of the parameters for the
which the model might be labeled as ``physical.''

The Cu site Coulomb repulsion $U_d$ is a large energy, making doubly
occupancy of Cu sites by two holes very unfavorable. The next largest
parameter, the charge transfer energy $\epsilon =
\epsilon_p-\epsilon_d>0$, plays a special role. Dependent on the
relative values of $\epsilon$, $U_d$ and the bandwidth $W$, the system
can be classified in different regimes:\cite{zaanen} the
charge-transfer regime with $U_d>\epsilon>W$ or the mixed-valent
regime with $U_d > W > \epsilon$, where $W$ is some measure of the
width of the lower band. Estimates place the cuprate superconductors
in the charge-transfer regime. The role of the Cu-O hybridization
$t_{pd}$ is important. Through the super-exchange mechanism, this
hybdization generates an antiferromagnetic exchange interaction
between the spins in the Cu sites. The O-O hybridization $t_{pp}$ and
the O site Coulomb repulsion $U_p$ are the two smallest
energies. 
%%%Somewhat arbitrarily we kept only $t_{pd}$. 
When $t_{pp}$ is non-zero, the non-interacting band structure has 
features that seem to appear in the normal state properties of the 
cuprate materials. We were mainly interested is studying the 
consequences of $V_{pd}$. These consequences were studied in part 
by previous QMC simulations of Dopf et al. \cite{dopf2} and Scalettar 
et al. \cite{scalettar2} In what follows we will 
scale all the energies by $t_{pd}$.

For the non-interacting case ($U_p=U_p=V_{pd}=0$), the band structure
is easily determined numerically and is illustrated in Fig.~2 for a
set of parameters used.  We will add and remove holes from lower band,
and this band is said to be half-filled when there is one hole per
unit cell.  At half-filling, for a wide range of parameters, the
ground-state is an anti-ferromagnetic insulator just like the one-band
Hubbard model. This band gap can be estimated from Fig.~2. A better
physical feel can be obtained from the exact expression easily
obtained if $t_{pp}=0$: It is simply $\epsilon$. The band-width $W=
\sqrt{(\epsilon/2)^2 +8}-\epsilon/2$. It varies
monotonically from 8 to zero as $\epsilon$ varies from zero to
infinity. $\epsilon=2$ marks a value for which $W=\epsilon$. This
picture does not change much for relatively small non-zero values of
$t_{pp}$.
As shown by Dopf et al. \cite{dopf2}, for $\epsilon=1$ the charge transfer gap
is vanishing small, whereas for $\epsilon=3$ a finite charge transfer gap
arises in the strong-coupling region. $\epsilon=1$ is in the mixed valent
regime, while $\epsilon=3$ is in the charge transfer regime. 
And we will see that $\epsilon=2$ behaves more like the charge-transfer 
than the mixed-valent regime. In Fig.~3, we show the Fermi surfaces for
infinite systems at the various dopings studied.

If $\epsilon \gg U_d $, the three-band model maps into a one-band
model with $t_{eff} \sim t_{pd}^2/ \epsilon$ and $U = U_d $. For $U_d
\gg t_{eff}$, the one-band model can in turn be mapped into the t-J
model with $J = 4t_{eff}^2/U_d$. Zhang and Rice \cite{zhang} have
argued that the t-J model can also be appropriate when $0< t_{pd} \ll
\epsilon$, $U_d$, $U_d - \epsilon$. In real materials, $\epsilon/t_{pd}$
is estimated to be $ \sim 2.7-3.7 $.\cite{martin} Therefore, besides
the lack of conclusive evidence that the one-band model superconducts,
it is also unclear that the mapping among the most studied models is
appropriate for physical values of the parameters.

With the numerical method used, a variety of expectation values can be
computed.  We focused on the pairing, spin, and charge correlation
functions.  More specifically we computed the extended s-wave and the
$d_{x^2-y^2}$ pairing correlations as functions of distance
\begin{equation}
P_\alpha(\vec R) = \langle\Delta_\alpha^\dagger(\vec R) \Delta_\alpha(0)\rangle
\end{equation}
where
\begin{eqnarray*}
\Delta_\alpha(\vec R) = \sum\limits_{\vec{\delta}}
f_\alpha(\vec{\delta}) \{&[&d_{\vec{R}\uparrow}d_{\vec{R}+\vec{\delta}
\downarrow} -d_{\vec{R}\downarrow}d_{\vec{R}+\vec{\delta}\uparrow}]\\
+ &[&p^x_{\vec{R}\uparrow}p^x_{\vec{R}+\vec{\delta}\downarrow}
-p^x_{\vec{R}\downarrow}p^x_{\vec{R}+\vec{\delta}\uparrow} ]\\
+&[&p^y_{\vec{R}\uparrow}p^y_{\vec{R}+\vec{\delta}\downarrow}
-p^y_{\vec{R}\downarrow}p^y_{\vec{R}+\vec{\delta}\uparrow}] \}
\end{eqnarray*}
with $\vec{\delta} = \pm \hat{x}, \pm \hat{y}$.  For the extended s-wave
pairing $ f_{s^*}(\vec{\delta}) = 1 $ for all $\vec{\delta}$ and for
the $d_{x^2 - y^2}$ pairing, $f_{d}(\vec{\delta}) = 1 $ for
$\vec{\delta} = \pm \hat{x}$ and $f_{d}(\vec{\delta}) = -1 $ for
$\vec{\delta} = \pm \hat{y} $. The magnitude of these quantities are
dominated by a large peak in $P_\alpha(\vec R)$ when $R=|\vec R|$ is
less than a few nearest neighbor distances. Over these distances,
$P_\alpha$ measures local correlations among spin and charge, has
little information about long-range pairing correlations, and may give
a ``false positive'' indication of enhanced pairing.  Because of this
we will report neither the $q=0$ spatial Fourier transformation nor the
partial sums like $S_\alpha(L)= \sum_{R\le L}P_\alpha(\vec R)$ as done
in some previous works.\cite{scalettar1}$^{-}$\cite{kuroki} Instead we
will report $S_\alpha(L)= \sum_{R\ge L}P_\alpha(\vec R)$ where $L$ is
about two lattice spacings. We will also report the ``vertex
contribution'' to the correlation functions
(see, for example, White et. al. \cite{white1}) defined as follows:
\begin{equation}
V_\alpha(\vec R) = P_\alpha(\vec R) - \bar{P}_\alpha(\vec R) ~,
\end{equation}
where $\bar{P}_\alpha(\vec R)$ is the contribution of dressed
non-interacting propagator: for each term in $ P_\alpha(\vec R)$ of
the form $< c^\dagger_\uparrow c_\uparrow c^\dagger_\downarrow
c_\downarrow>$, $\bar{P}_\alpha(\vec R)$ has a term like $<
c^\dagger_\uparrow c_\uparrow><c^\dagger_\downarrow c_\downarrow>$. 
We found that in most cases the conclusions remain the same
no matter which quantity we look at.

For the static spin-spin correlation function we used the Fourier
transform of the spin-spin correlation function for the spin on the Cu
sites
\begin{equation}
S(k) = \frac{1}{N}\sum_{lm} e^{ik\cdot(l-m)}
   \langle (n_{l,\uparrow}^d-n_{l,\downarrow}^d)
           (n_{l+m,\uparrow}^d-n_{l+m,\downarrow}^d)\rangle ~,
\end{equation}
where $l$ and $m$ refers to the Cu sites and $N$ is the number of unit
cells. For charge-charge correlations we computed a charge-transfer
correlation function involving the O sites neighboring a Cu site
quantity:
\begin{equation}
C(k)=\frac{1}{N}\sum_{ij}e^{ik\cdot(i-j)}  \langle\rho(i)\rho(j)\rangle.
\end{equation}
where $j$ are the nearest-neighbors of $i$,
$\rho(i)=n_{i}^{d}-n_{i}^{p_x}-n_{i}^{p_y}$ with $n_{i}^{d}$,
$n_{i}^{p_x}$, and $n_{i}^{p_y}$ being the charge-density operators on
the Cu, x-axis O, and y-axis O in the unit cell $i$.

\section{\label{CPMC} Numerical Method}

Our numerical method, the constrained path Monte Carlo (CPMC) method,
is extensively described and benchmarked
elsewhere \cite{zhang1,zhang2}. Here we only discuss its basic strategy
and approximation.  In the CPMC method, the ground-state wave function
$|\psi_0\rangle$ is projected from a known initial wave function
$|\psi_T\rangle$ by a branching random walk in an over-complete space
of Slater determinants $|\phi\rangle$.  In such a space, we can write
$|\psi_0\rangle = \sum_\phi \chi(\phi) |\phi\rangle$.  The random walk
produces an ensemble of $|\phi\rangle$, called random walkers, which
represent $|\psi_0\rangle$ in the sense that their distribution is a
Monte Carlo sampling of $\chi(\phi)$, that is, a sampling of the
ground-state wave function. More specifically, starting with some
trial state $|\psi_T\rangle$, we project out the ground state by
iterating
\begin{equation}
|\psi'\rangle = e^{-\Delta\tau (H-E_T)}|\psi\rangle
\end{equation}
where $E_T$ is some guess of the ground-state energy.  Purposely
$\Delta\tau$ is a small parameter so for $H=T+V$ we can write
\begin{equation}
 e^{-\Delta\tau H}\approx e^{-\Delta\tau T/2}
                          e^{-\Delta\tau V}
                          e^{-\Delta\tau T/2}
\end{equation}
where $T$ and $V$ are the kinetic and potential energies.

For the study at hand, the initial state $|\psi_T\rangle$ is the
direct product of two spin Slater determinants, i.e.,
\begin{equation}
 |\psi_T\rangle = \prod_\sigma |\phi_T^\sigma\rangle
\end{equation}
Because the kinetic energy is a quadratic form in the creation and
destruction operators for each spin, the action of its exponential on
the trial state is simply to transform one direct product of Slater
determinants into another.  While the potential energy is not a
quadratic form in the creation and destruction operators, its
exponential is replaced by sum of exponentials of such forms via the
discrete Hubbard-Stratonovich transformation. For the on-site Coulomb
term, this transformation is
\begin{eqnarray}
e^{-\Delta\tau U_d n_{i,\sigma}^d n_{i,-\sigma}^d}
&=& \frac{1}{2} \sum_{x=\pm 1}e^{-x \Delta\tau
J_d(n_{i,\sigma}^d-n_{i,-\sigma}^d)}\nonumber\\
&&e^{\frac{1}{2}\Delta\tau
U_d(n_{i,\sigma}^d+n_{i,-\sigma}^d)}
\end{eqnarray}
provided $U_d\ge 0$ and $\cosh \Delta\tau J_d = e^{-\Delta\tau
U_d/2}$. For the nearest neighbor Coulomb repulsion term, we make the
same type of transformation but we have to do it many more times:
$n_{i}^d n_{j}^p=n_{i\uparrow}^d n_{j\uparrow}^p+n_{i\uparrow}^d
n_{j\downarrow}^p+n_{i\downarrow}^d n_{j\uparrow}^p+n_{i\downarrow}^d
n_{j\downarrow}^p$. For each term and each $j$ in the $\hat x$ and
$\hat y$ directions, a Hubbard-Stratonovich transformation is required
for a total of 8 such transformations. Because the computational time
scales with the number of Hubbard-Stratonovich transformations, having
a $V_{pd}\not=0$ increases the computational cost by a factor of 8.

One consequence of the Hubbard-Stratonvich transformation is the
factorization of the projection into an up and down spin
part. Accordingly we re-express the iteration step as
\begin{equation}
 \prod_\sigma |\phi_\sigma'\rangle = \int d\vec x\, P(\vec x)
    \prod_\sigma B_\sigma(\vec x)|\phi_\sigma\rangle
\end{equation}
where $\vec x =(x_1,x_2,\dots,x_N)$ is the set of Hubbard-Stratonovich
fields (one for each lattice site), $N$ is the number of lattice
sites, $P(\vec x)=(\frac{1}{2})^N$ is the probability distribution for
these fields, and $B_\sigma(\vec x)$ is an operator function of these
fields formed from the product of the exponentials of the kinetic and
potential energies.

The Monte Carlo method is used to perform the multi-dimensional
integration over the Hubbard-Stratonovich fields. It does so by
generating a set of random walkers initialized by replicating
$|\psi_T\rangle$ many times. Each walker is then propagated
independently by sampling a $\vec x$ from $P(\vec x)$ and propagating
it with $B(\vec x)$. After the propagation has ``equilibrated,'' the
sum over the walkers provides an estimate of the ground-state wave
function $|\psi_0\rangle$.

We used two different estimators for the expectation values of some observable
${\cal O}$. One is the mixed estimator
\begin{equation}
 \langle {\cal O}\rangle_{\mathrm{mixed}} =
     \frac{\langle\psi_T|{\cal O}|\psi_0\rangle}
          {\langle\psi_T|\psi_0\rangle}
\end{equation}
and the other is the back-propagated estimator
\begin{equation}
 \langle{\cal O}\rangle_{\mathrm{bp}} =
     \frac{\langle\psi_T|e^{-\ell\Delta\tau H}{\cal O}|\psi_0\rangle}
          {\langle\psi_T|e^{-\ell\Delta\tau H}|\psi_0\rangle}
\end{equation}
where $|\psi_0\rangle$ is the QMC estimate of the ground state and
$\ell$ is typically in the range of 20 to 40.  For observables that
commute with the Hamiltonian, the mixed estimator is a very accurate
one and converges to the exact answer as $|\psi_0\rangle$ converges to
exact ground state. For observables that do not commute with the
Hamiltonian, like correlation functions, the back-propagated estimator
has been found to give very accurate estimates of ground-state
properties. Significant differences between the predictions of these
two estimators often exist.

To completely specify the ground-state wave function for a system of
interacting electrons, only determinants satisfying $\langle
\psi_0|\phi\rangle>0$ are needed because $|\psi_0\rangle$ resides in
either of two degenerate halves of the Slater determinant space,
separated by a nodal surface ${\bf N}$ that is defined by $\langle
\psi_0|\phi\rangle=0$.  The degeneracy is a consequence of both
$|\psi_0\rangle$ and $-|\psi_0\rangle$ satisfying Schr\"odinger's
equation.  The sign problem occurs because walkers can cross ${\bf N}$
as their orbitals evolve continuously in the random
walk. Asymptotically they populate the two halves equally, leading to
an ensemble that has zero overlap with $|\psi_0\rangle$.  If ${\bf N}$
were known, we would simply constrain the random walk to one half of
the space and obtain an exact solution of Schr\"odinger's equation. In
the constrained-path QMC method, without {\it a priori\/} knowledge of
${\bf N}$, we use a trial wave function $|\psi_T\rangle$ and require
$\langle \psi_T|\phi\rangle>0$.  This is what is called the
constrained-path approximation.

All the calculations reported here were done for copper-oxide planes
with periodic boundary conditions.  Mostly, we study closed shell
cases, for which the corresponding free-electron wave function is
non-degenerate and translationally invariant. For $6\times 6$ unit
cells, the dopings, producing Fermi surfaces in Fig.~2, correspond to
closed shell fillings. In these cases, the free-electron wave
function, represented by a single Slater determinant, is used as the
trial wave function $|\psi_T\rangle $. The use of an unrestricted
Hartree-Fock wave function as $|\psi_T\rangle $ generally produced no
significant improvement in the results. At half-filing, which is not a
closed shell case, we used a linear combination of two degenerate
${\bf Q}=(\pi,\pi)$ spin-density wave states.

In a typical run, the average number of walkers was 600, and the time
step was 0.03.  We performed 1600 steps before we started taking
measurements, and we did the measurements in 30 blocks of 320 steps
each to ensure statistical independence. Back propagation measurements
had 40 backward steps.

\section{\label{RESULTS} Results}

As mentioned before, all our simulations were done on lattices of
$6\times 6$ unit cells. For this size, 36 holes corresponds to a 
half-filed case.
In units of $t_{pd}$, we set $U_d=6$, $U_p=0$, and $t_{pp}=0.3$
for most studied cases.
We varied $V_{pd}$ between 0 and 1 for several different hole fillings 
and values of the charge-transfer energy $\epsilon$.
We were mainly concerned with hole doped cases. % so we increased the 
number of holes.

\subsection{\label{CHARGE} Charge Correlation Functions}
 
In Fig.~4 we show the expectation values of the charge on the Cu sites
as a function of $V_{pd}$ for several band fillings and values of the
charge-transfer energy $\epsilon$.
When $V_{pd}=0$, we see that for
the half-filled case, even with a relatively large value of
$\epsilon$, there are substantial holes distributed on the O sites.
When doped to 42 and 46 holes, most of the added
holes go to the O site. Except for the $\epsilon=1$ case, increasing
$V_{pd}$ from 0 transfers some of the O charge to the Cu. The transfer
rate increases if $\epsilon$ is increased. These are expected results:
in the charge-transfer regime, $U_d>\epsilon>W$, with a repulsive
$V_{pd}$, it become energetically favorable for some charge to move
from O to Cu even at the expense of some unfavorable double occupancy
of the Cu site caused by a large $U_d$.  On the other hand, in the
mixed-valent regime, $U_d>W>\epsilon$, when $\epsilon=1$, we see a
movement of holes from the Cu site to the O sites. Here the energy
difference between the Cu and O states is smaller, and a strong
on-site repulsive $U_d$ favoring charge removable from the Cu sites
dominates the smaller repulsive $V_{pd}$, opposing the movement of
charge to the O sites. In general, the presence of $V_{pd}$ seems to
expand the charge-transfer regime. Similar results have been seen in
finite-temperature QMC \cite{dopf2} and zero-temperature exact 
diagonalization \cite{scalettar2} studies.

Another effect of increasing $V_{pd}$ is the decreasing of the correlation
between the charge on the Cu and the neighboring O sites. This is
shown in Fig.~5. At a given band filling, increasing $\epsilon$ only
has a little effect. Similar behavior was also seen in a
finite-temperature QMC study.
One also observes that in the charge-transfer regime the decreasing rate
seems independent of the filling and $\epsilon$.
Because $C(0)\approx N_h/N - \langle n_{Cu}n_O\rangle$, it is no surprise to
observe an increase in $C({\bf k}=0)$ with an increasing $V_{pd}$.

It is instructive to compare our findings with Stephan et al.'s \cite{stephan}
results. Their exact diagonalization results showed
that when $U_{Cu}=U_{O}=\infty$, doped holes make the hole on the Cu
sites transfer to the O sites, and with increasing $V_{pd}$, the charge
on the Cu site decreases continuously. They also found that the addition
of a second hole produces a smaller additional change in the neighboring
Cu-O charge correlation than the first one does. Both behaviors of
the hole on the Cu site and neighboring Cu-O charge correlation indicate that
a charge-transfer ``bipolaron'' forms in this system. Therefore 
they concluded that the binding energy is obtained from electronic
polarization. From our simulation results, in the charge transfer region,
the charge on the Cu site increases with hole doping and increasing $V_{pd}$,
suggesting that electronic polarization has no effect in 
the physically relevant region. Hence we expect that the binding energy
is mainly gained from magnetic mechanism.

\subsection{\label{SPIN} Spin Correlation Functions}

In Fig.~6 we show the behavior of the local magnetic moment on the Cu
sites. First, we see that increasing the hole doping increases the
moment. More specifically, at a given doping, increasing $\epsilon$
increases the moment. When $\epsilon > 1$, increasing $V_{pd}$
increases the moment, but when $\epsilon = 1$, increasing $V_{pd}$
decreases it. Clearly, the increase of the moment is strongly
correlated with the increase and decrease of charges on the Cu sites.

In Figs.~7 and 8 we show the wavevector dependence of the Fourier
transform of the static spin-spin correlation function for the Cu
sites as a function of $V_{pd}$ for a doping to 42 holes.  This
function is plotted along high symmetry lines in the first Brillouin
zone. The different figures corresponds to different values of
$\epsilon$. In mixed valent regime, Fig.~7, we see that
increasing $V_{pd}$ suppress this function over the entire zone. This
suppression is consistent with the suppression of the local moment seen
in Fig.~6. On the other hand, in the charge-transfer regime, Fig.~8,
increasing $V_{pd}$ enhances this function, again consistent
with the enhancement of the local moment seen in Fig.~6. By comparing
the two figures we see that for a given $V_{pd}$ increasing $\epsilon$
increases these correlations and sharpens the peaks in the
functions. In each figure there are two principal peaks: One is
connected with the displacement of the antiferromagnetic peak to
$(\pi,\pi-\delta)$. The other is the appearance of an incommensurate
structure at $(\pi-\delta',\pi-\delta')$. A weaker spin-density wave
structure is at $(\pi,0)$.

Previous QMC simulations of the one-band Hubbard model \cite
{moreo1,moreo2} have seen a similar shifting (and splitting) of the 
peak in the static structure factor from the antiferromagnetic position 
$(\pi,\pi)$ to positions $(\pi,\pi-\delta)$ on the face of the Brillouin 
zone and $(\pi-\delta',\pi-\delta')$ along the diagonal direction. This 
behavior is in agreement with the experimental data for LSCO presented
in Fig.~3 of Ref.\cite{mason}, where a minimum is observed at $(\pi,\pi)$
along the diagonal direction. Our CPMC simulations of $t-t^{'}-U$ Hubbard
model also found that for a large $t^{'}$, a weak peak appears along the 
diagonal direction.
%We are unaware on any previous observation of the incommensurate peak at
%$(\pi-\delta',\pi-\delta')$ in a QMC simulation of either the one-band
%or three-band Hubbard models. 
We remark that we did not study the
dependence of either of these peaks on lattice size.% and hence made no
%attempt to establish whether either or both represent a possible state
%of long-range order.

To examine whether the incommensurate peak along the diagonal 
direction is produced by a finite O-O hopping $t_{pp}$, in Fig.~9 (a)
and (b) we display the spin structure factor $S(k)$ as a function of 
$t_{pp}$ for different $\epsilon$. The parameters are the same as in 
Fig.~7. Here a nonzero $t_{pp}$ makes the hole filling correspond to a
closed-shell case. From Fig.~9 (b), even for a very small $t_{pp}$, 
a weak peak at the $(\pi-\delta',\pi-\delta')$ still exists.
%Hence the intrinsic physics of cuprates compounds can be described in
%the three-band Hubbard model with three important parameters:
%$t_{pd}, U_{d}$, and $\epsilon$.
With increasing $t_{pp}$, the spin-spin
correlations are strongly suppressed near the AF wavevector $(\pi,\pi)$,
and at the same time the amplitude of the incommensurate peak along the 
diagonal direction or tendency to this peak forming is enhanced.
For the half-filling case (data not shown) we also observed that
increasing $t_{pp}$ 
greatly suppresses AF order.

Finally, we report how the AF long range order in the half-filling
case is destroyed by the hole doping. In Fig.~10(a) and (b), the spin structure
factor is plotted for different hole filling cases. From Fig.~10(a), it
is clearly seen that AF spin-spin correlation is strongly suppressed
by the hole doping. As shown in Fig.~10(b), in the light hole doping 
region ($N_h=42$), there exist two incommensurate peaks. When the system
is doped to $N_h=46$, the incommensurate peak along the diagonal direction
disappears, but the peak on the face of of the Brillouin zone is robust.
In the heavy doping region ($N_h=54$), the spin structure factor is
featureless near ($\pi,\pi$), and the peak occurs at ($\pi,0$).

\subsection{\label{PAIRING} Pairing Correlation Functions}

A typical pairing correlation function as a function of distance and
$V_{pd}$ is shown in Fig.~11. Here we show the d-wave function and see
that it is dominated by a large peak at short distances ($R<2$). At these
distances, increasing $V_{pd}$ increases the magnitudes of the
correlations slightly. At larger distances ($R>2$), the trend reverses.
The dominance of the local peak is such that a measure of pairing like 
the ${\bf k}=0$ dependence of the Fourier transform of the pairing
correlation function or the integral of $P(R)$ with a large distance
cut-off can exhibit behavior indicative of the only the short-range
behavior and hide the more relevant long-range behavior. 

The short-range behavior is illustrated in more detail in Fig.~12. For
two different values of $U_d$, we plot the value of the $R=0$ peak in
the vertex contribution for both extended s-wave and d-wave symmetries
for several values of $\epsilon$. In both the charge-transfer and
mixed valent regimes, the $R=0$ value for both symmetries increases
monotonically with increasing $V_{pd}$.

For comparison, the long-range ($R>2$) behavior is illustrated in
Fig.~13. In both the s and d-wave channels, the long-range vertex
contributions in the charge-transfer regime decrease with increasing
$V_{pd}$ but in the mixed-valent regime it increases.  Over the range
of $V_{pd}$ simulated, the long-range part of the d-wave contribution is
consistently larger than the s-wave contribution.

As a function of filling, the behavior is more complex. In Fig.~14 is
the short-range part of the vertex contribution. In the mixed valent
regime, both the extended s-wave and d-wave channels decrease rapidly
with electron and hole doping. In the charge-transfer regime, the
d-wave functions falls with doping but the extended s-wave
contribution initially increases. The long-range contribution, shown
in Fig~15, shows the dominance of the d-wave channel. In the
mixed-valent regime it basically decreases with doping, whereas in the
charge-transfer regime, it initially increases but only to decrease
rapidly for large hole doping. Our findings suggest that the
conclusions of Dopf et al.\cite{dopf2} reflect the
behavior of local interaction vertex, not the long range property.

\section{\label{CONCLUSIONS} Summary and Conclusions}

We summarize our results as follows: Using the CPMC method, we
simulated the two-dimensional three-band Hubbard model to study its
charge, spin, and pairing correlations as a function of electron and
hole doping, and the charge-transfer energy $\epsilon$ and the Coulomb
repulsion $V_{pd}$ between charges on neighboring Cu and O lattice
sites. We found that increasing $V_{pd}$ suppressed the charge-charge
correlations between neighboring Cu and O sites. In the mixed-valent
regime it had the effect moving small amounts of charge from the Cu
sites to the O sites. In the charge-transfer regime, the effect was
the opposite. Upon hole doping, more of the extra holes went to the O
sites than to the Cu sites.
 
At a given doping, increasing $V_{pd}$ increased the spin-spin
correlations in the charge-transfer regime but decreased them in the
mixed-valent regime. Also electron and hole doping away from
half-filling was accompanied by a rapid suppression of
anti-ferromagnetic correlations. As a function of doping, $\epsilon$,
and $V_{pd}$, the behavior of the magnetic moment on the Cu sites was
strongly correlated with the behavior of the charge on the Cu sites.

As a function of distance, both the $d_{x^2 - y^2}$-wave and extended
s-wave pairing correlations decayed quickly. In the charge-transfer
regime, increasing $V_{pd}$ decreased the long-range part of the
correlation functions in both channels, while in the mixed-valent
regime, it increased the long-range part of the s-wave behavior but
decreased that of the d-wave behavior decreased. Still the d-wave
behavior dominated.  At a given doping, increasing $V_{pd}$ increased
the spin-spin correlations in the charge-transfer regime but decreased
them in the mixed-valent regime. Also electron and hole doping away
from half-filling was accompanied by a rapid suppression of
anti-ferromagnetic correlations.

We presented a more extensive study of the effects of $V_{pd}$ than
previous QMC studies at zero and finite temperature. Our results
illustrate the presence of both s and d-wave correlations in the
charge-transfer regime, with the d-wave correlations generally
dominating. These results highlight the difference in the behavior
between the short and long range part of these correlation
function. Figures of merits that include the short-range part are
dominated by the behavior of the short-range part. For the system size
studied the correlations are weak. We did not study these
correlations as function of system size. We believe the size dependence
will be the same as previous studies.

Lastly, as a complementary part to results presented so far, we briefly report
the effects of $U_p$ on charge, magnetic and pairing correlations. In both
the mixed valent and charge-transfer regions, and for all band fillings,
we found that a finite $U_p$ ($U_{p}=2.0$) moves some charge from the oxygen
sites to the copper sites, which causes an increasing of magnetic moment
at the copper sites. Consistent with previous observations \cite{hirsch,guerr}
that $U_p$ has a negative effect on the hole binding, the long-range part of 
the d-wave correlation functions is suppressed by $U_p$. Our simulation 
results also show that $U_p$ has a larger effect in the mixed valent region 
than that in the charge-transfer region.

\section*{Acknowledgments}
The work of Z. B. H. was supported in part by the
Earmarked Grant for Research from the Research Grants Council (RGC) of
the HKSAR under Projects CUHK 4190/97P-2160089.  The work of J. E. G.
was supported by the US Department of Energy. Part of his work was
performed while as a guest of the Chinese University of Hong Kong.
He gratefully acknowledges this hospitality.

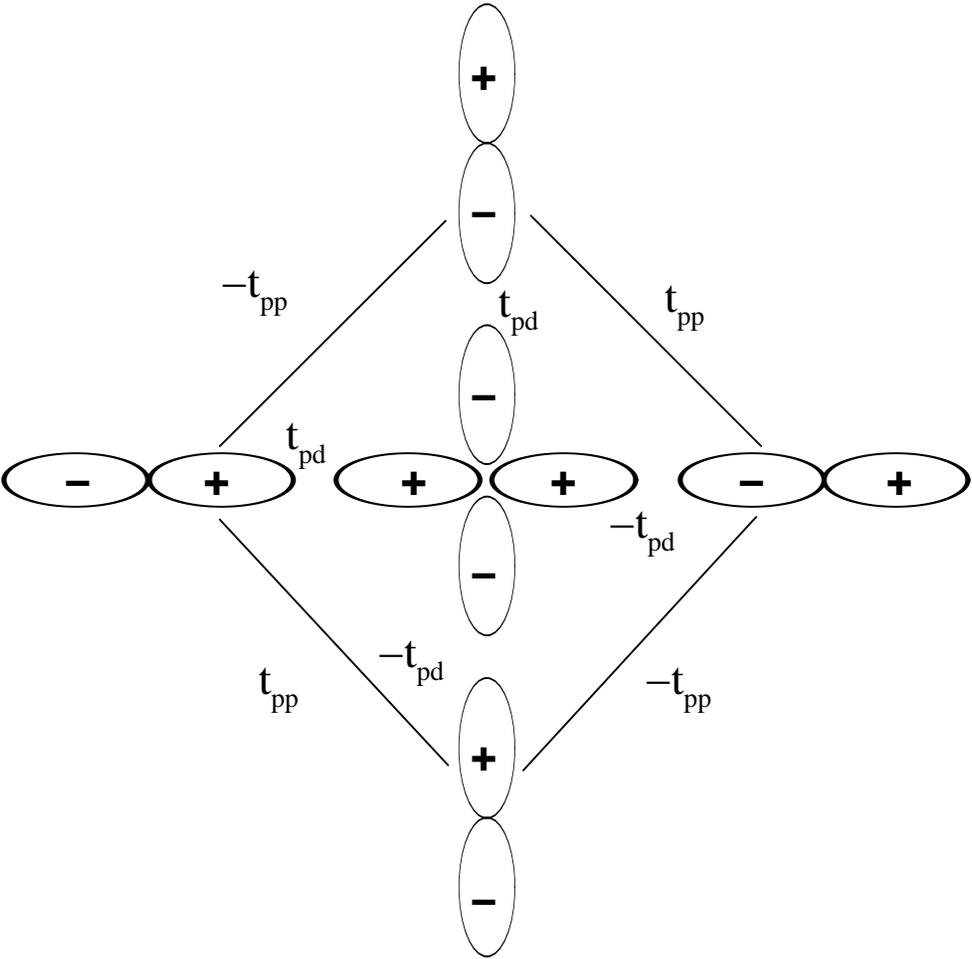
\begin{figure}[tbp]
\caption{Phase convention for the hopping matrix elements. The copper 
$d_{x^2-y^2}$
orbital is surrounded by the oxygen $p_x$ and $p_y$ orbitals. The hopping
matrix elements are shown with their corresponding phase.}
\end{figure}

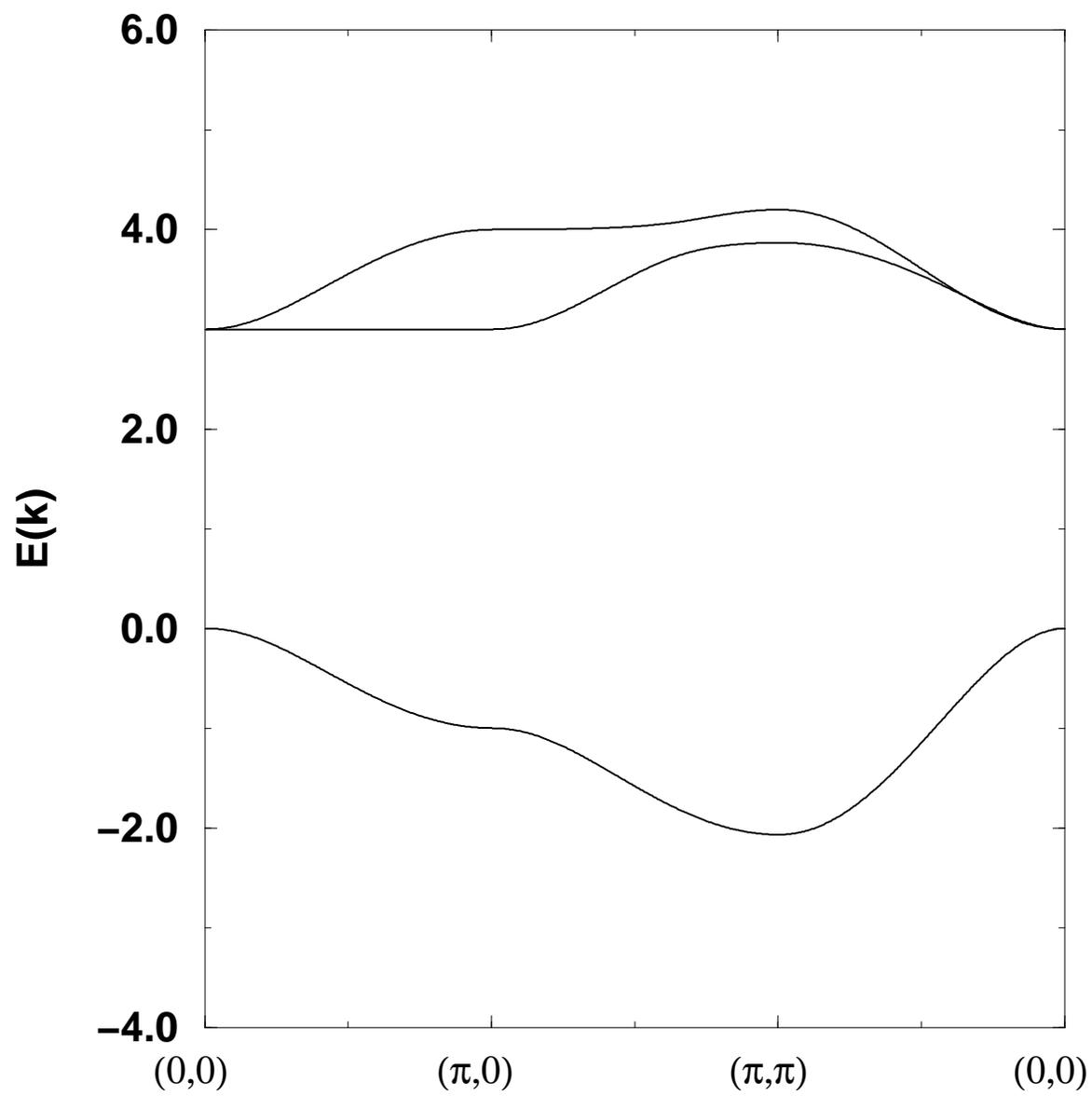
\begin{figure}[tbp]
\caption{The band-structure of an infinite systems for $\epsilon=3$
and $t_{pp}/t_{pd}=0.3$.}
\end{figure}

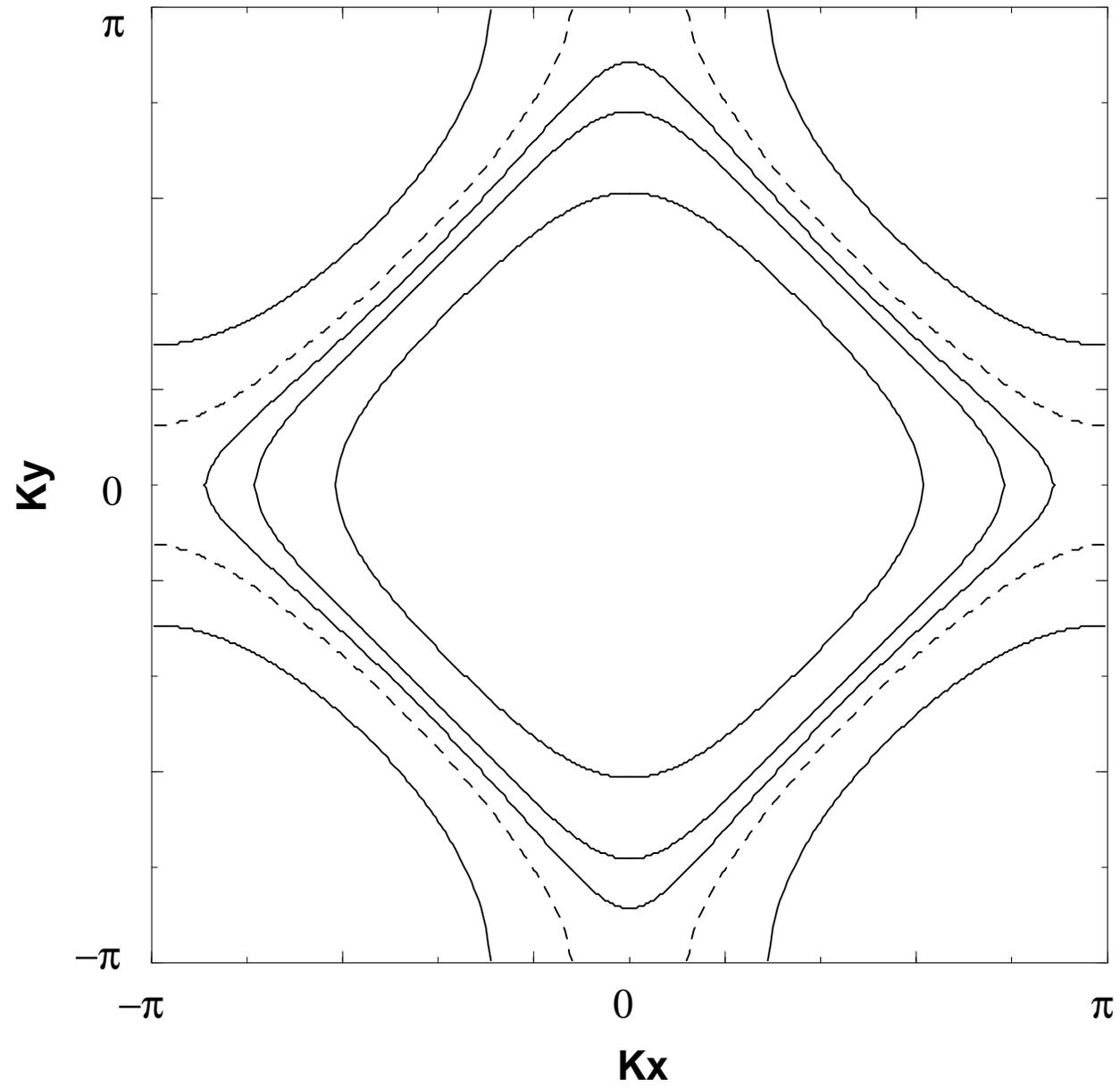
\begin{figure}[tbp]
\caption{Fermi surfaces of an infinite-system for $\epsilon=3$ and
$t_{pp}/t_{pd}=0.3$. From the inside out, the hole fillings are
54/36. 46/36, 42/36, 36/36 (dashed line), and 26/36.}
\end{figure}

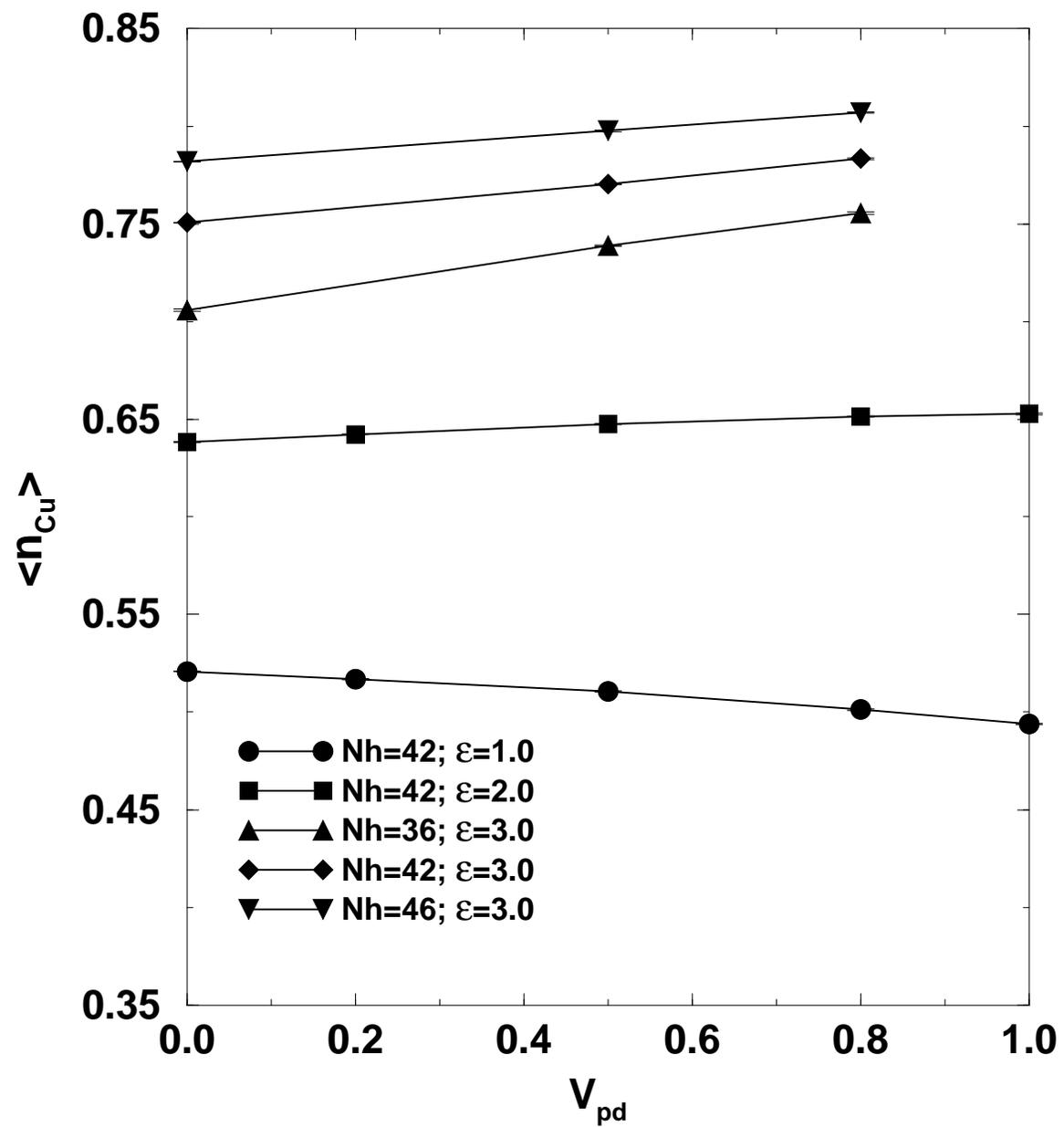
\begin{figure}[tbp]
\caption{Average charge on Cu sites as a function of $V_{pd}$ for different
fillings and charge-transfer energies $\epsilon$.}
\end{figure}

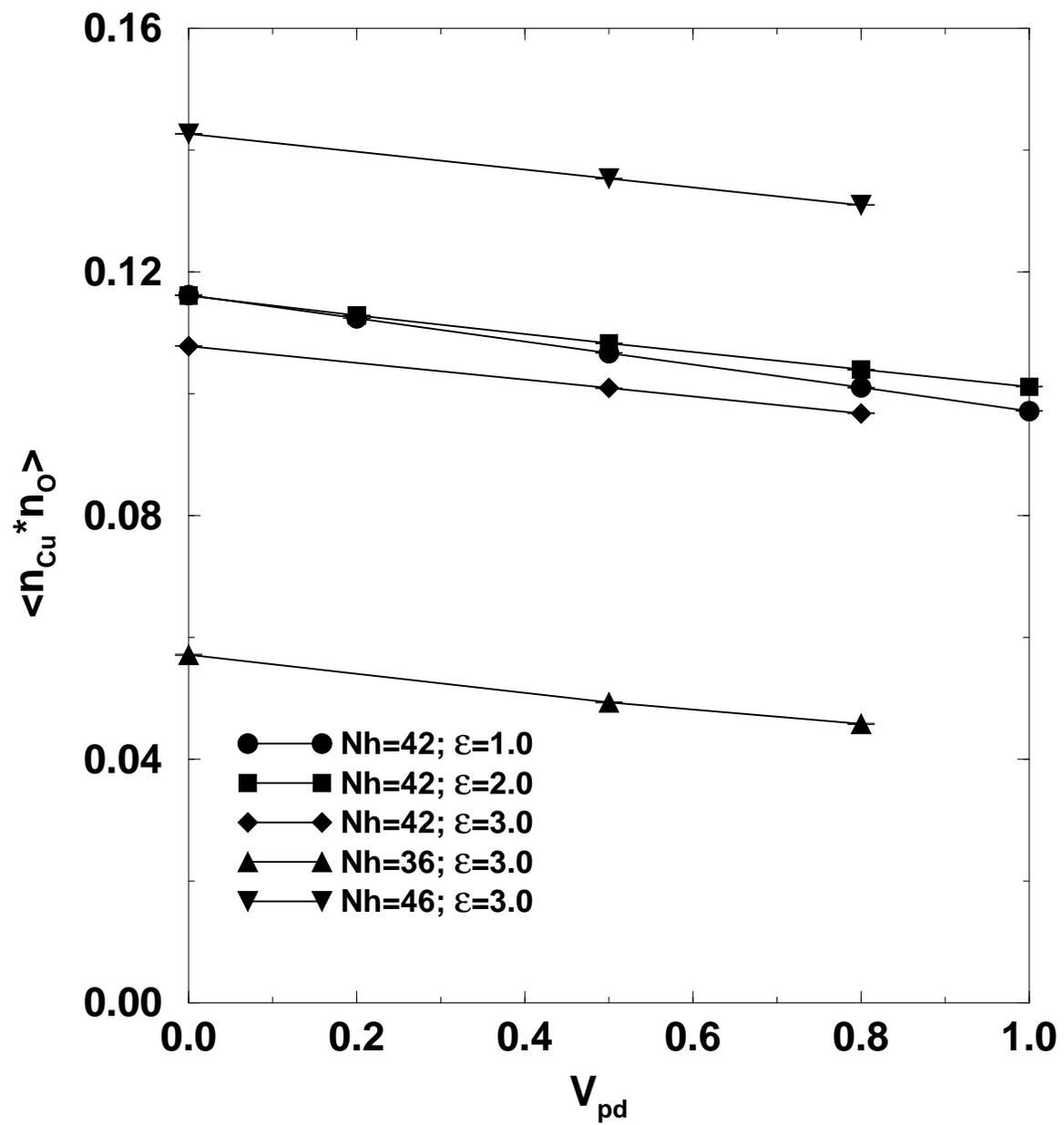
\begin{figure}[tbp]
\caption{Charge correlation between neighboring Cu and O sites as a function 
of $V_{pd}$ for different fillings and charge-transfer energies
$\epsilon$.}
\end{figure}

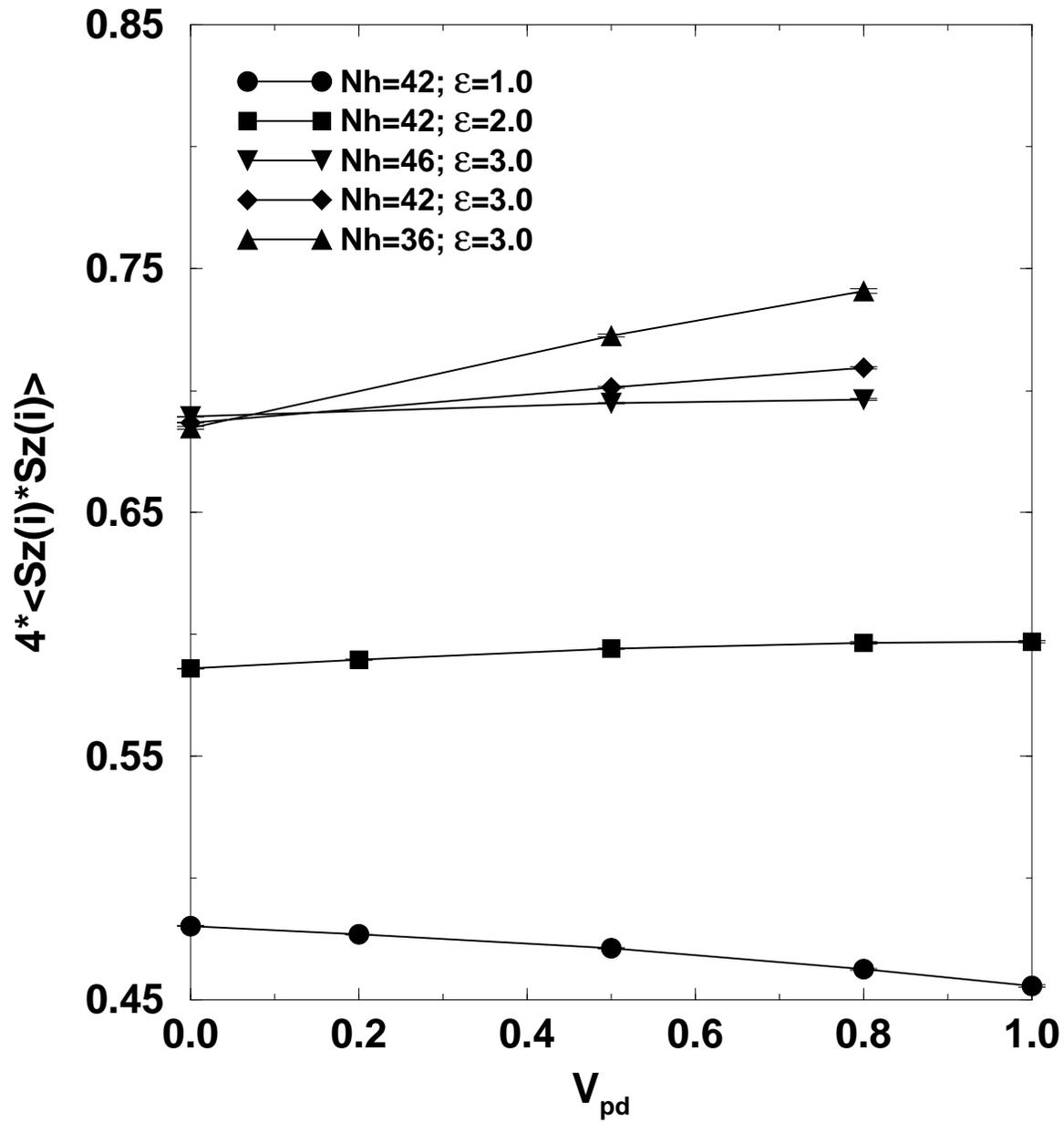
\begin{figure}[tbp]
\caption{Average magnetic moment at the Cu sites as a function of $V_{pd}$ for
different fillings and charge-transfer energies $\epsilon$.}
\end{figure}

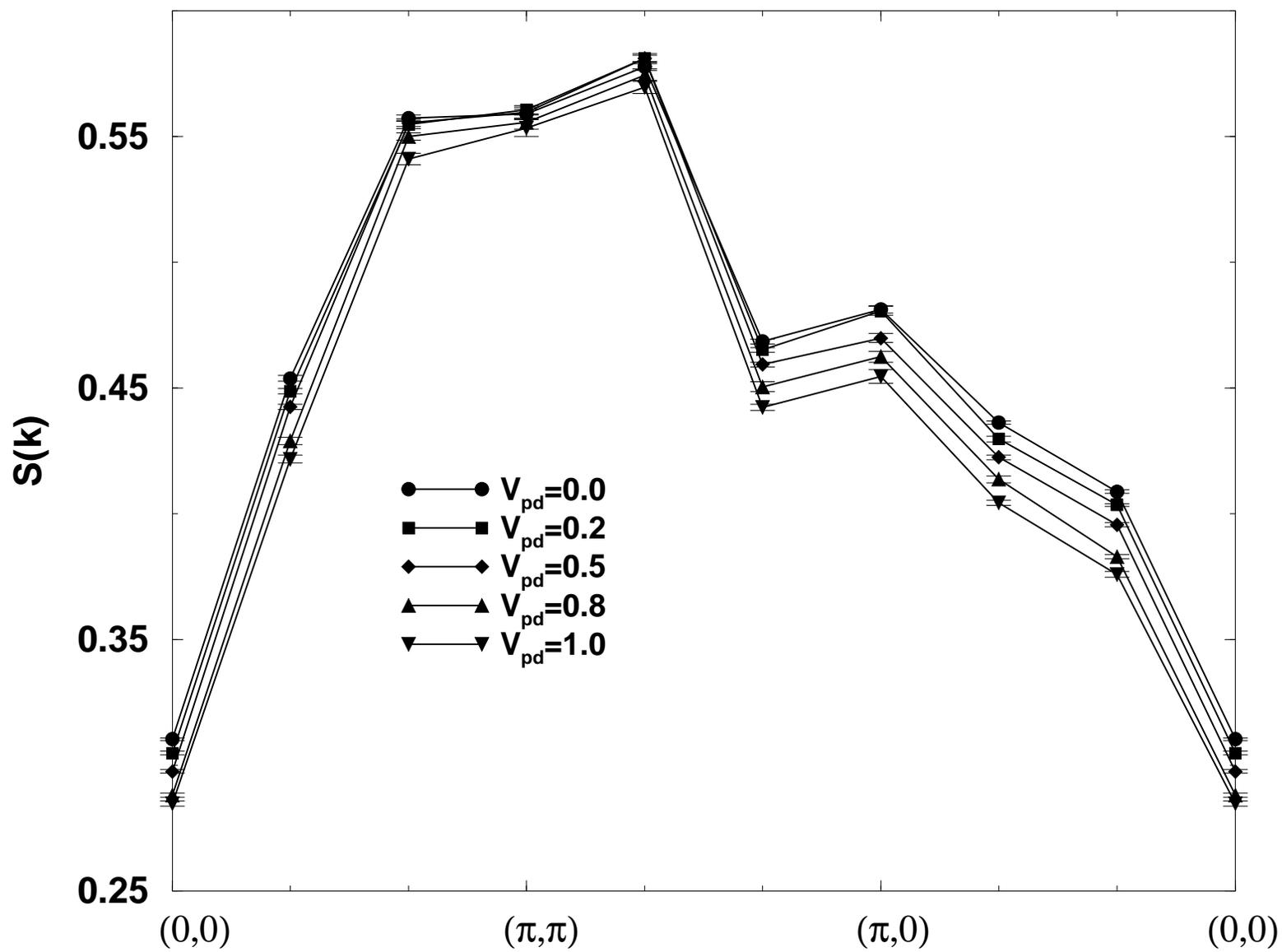
\begin{figure}[tbp]
\caption{Average static spin structure factor for Cu sites as a function of
the wavevector ${\bf k}$ and $V_{pd}$. $\epsilon=1$ and the number of
holes equals 42.}
\end{figure}

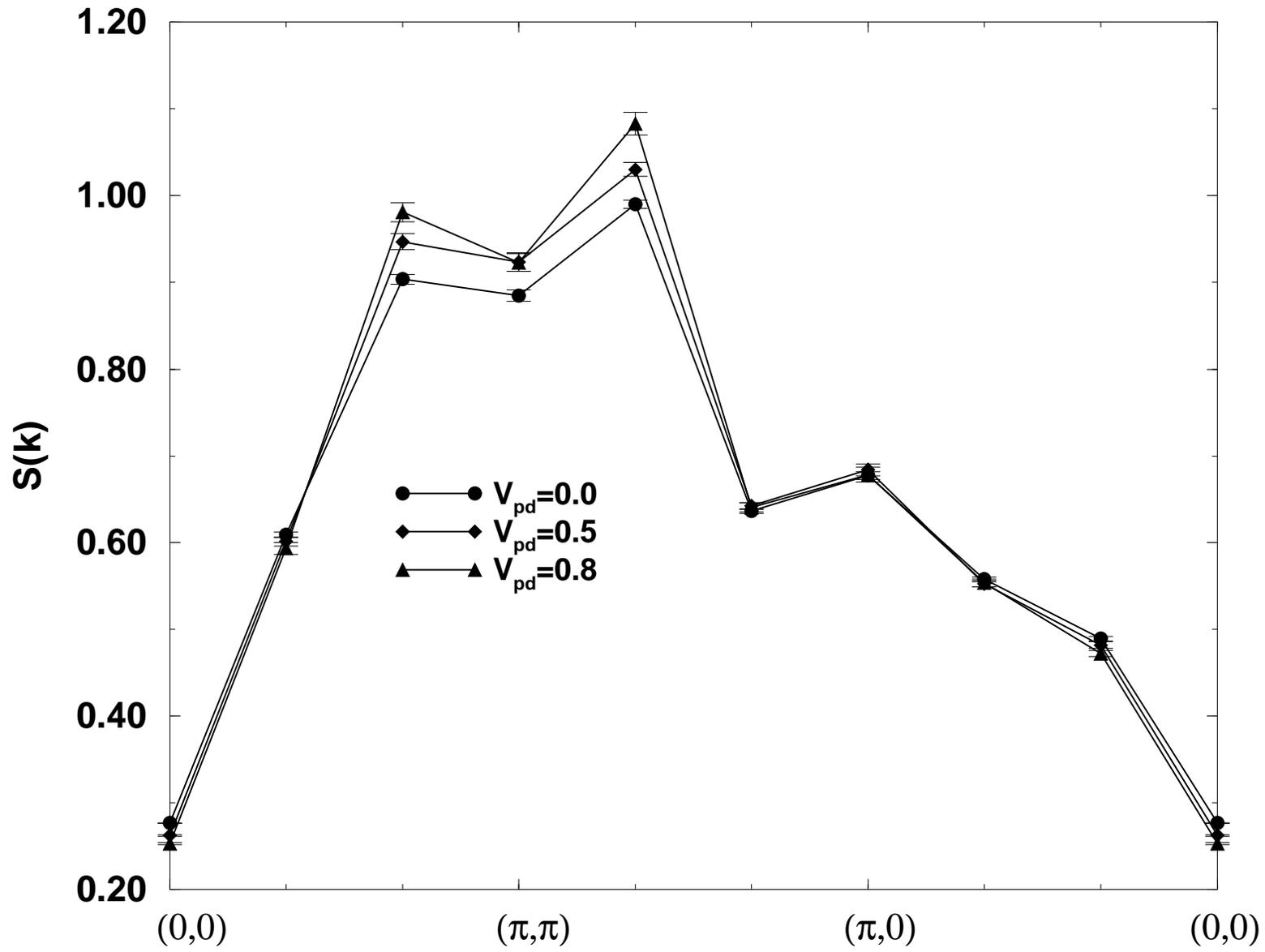
\begin{figure}[tbp]
\caption{Average static spin structure factor for Cu sites as a function of
the wavevector ${\bf k}$ and $V_{pd}$. $\epsilon=3$ and the number of
holes equals 42.}
\end{figure}

\begin{figure}[tbp]
\caption{Average static spin structure factor for Cu sites as a function of
the wavevector ${\bf k}$ and $t_{pp}$. (a) $\epsilon=1$, (b) $\epsilon=3$,
 and the number of holes equals 42.}
\end{figure}

\begin{figure}[tbp]
\caption{Average static spin structure factor for Cu sites as a function of
the wavevector ${\bf k}$ and $N_h$ with $\epsilon=3$.
(a) Half-filling and doped cases and (b) Doped cases.}
\end{figure}

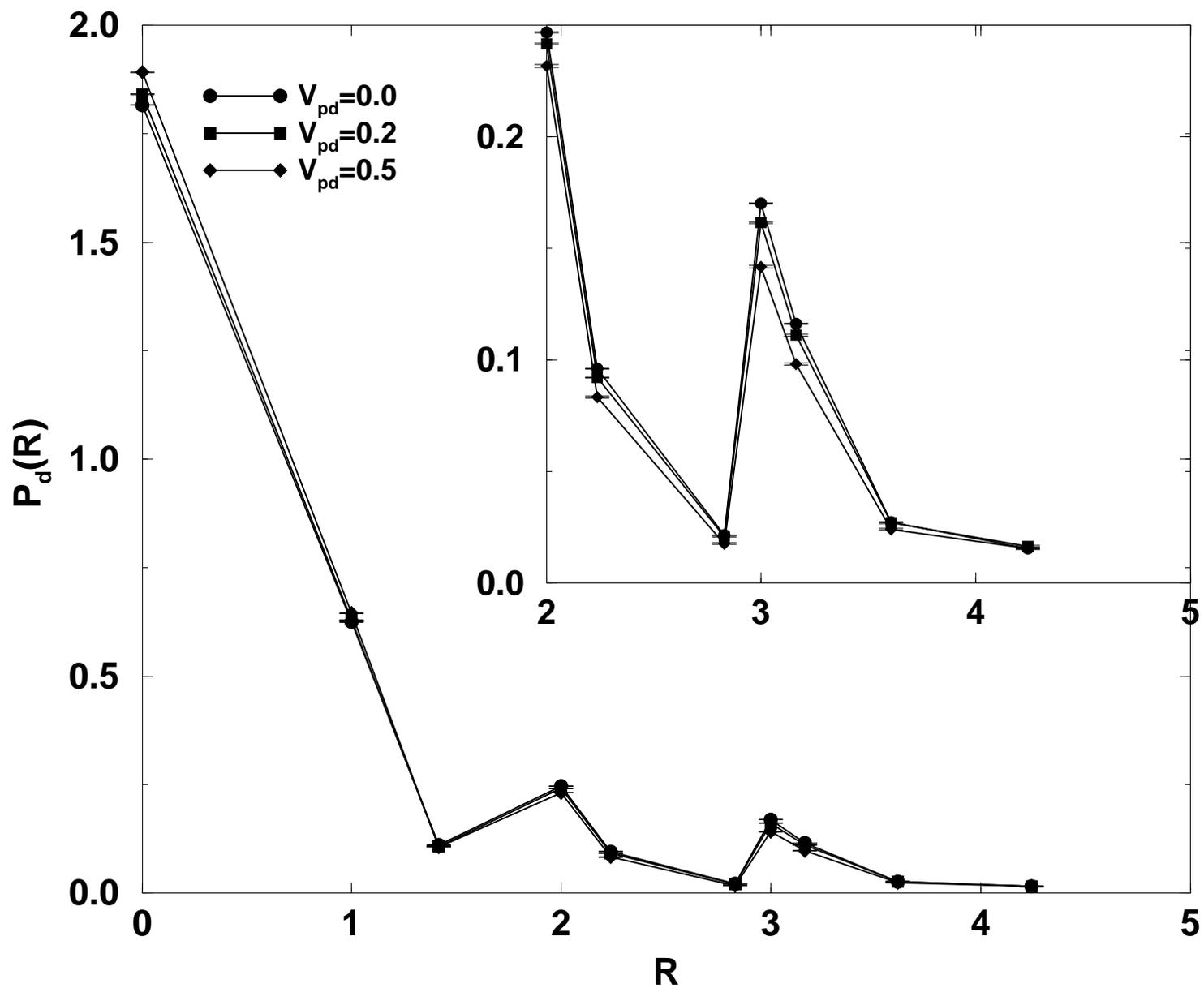
\begin{figure}[tbp]
\caption{d-wave pairing correlation function as a function of distance $R$ for
different  values of $V_{pd}$. The number of holes equals 42.}
\end{figure}

\begin{figure}[tbp]
\caption{Local ($R=0$) vertex contributions to the extended s and d-wave
pairing correlation function  as a function of $V_{pd}$ for different
values of $\epsilon$ and $V_{pd}$. The number of holes equals 42.}
\end{figure}

\begin{figure}[tbp]
\caption{Long-range (averaged for $R>2$) part of the vertex contributions
(averaged over $R>2$) to the extended s and d-wave pairing correlation
function as a function of $V_{pd}$ for different values of $\epsilon$
and $V_{pd}$. The number of holes equals 42.}
\end{figure}

\begin{figure}[tbp]
\caption{Local part ($R=0$) of the vertex contributions to the extended s and 
d-wave pairing correlation function as a function of filling. $V_{pd}=0$.The
number of holes equals 42. (a) $\epsilon=1$. (b) $\epsilon=3$.}
\end{figure}

\begin{figure}[tbp]
\caption{Long-range part (averaged for $R>2$) of the vertex contributions to
the extended s and d-wave pairing correlation function as a function
of filling. $V_{pd}=0$.The number of holes equals 42. (a)
$\epsilon=1$. (b) $\epsilon=3$.}
\end{figure}


\begin{thebibliography}{99}

\bibitem{white1} S. R. White, D. J. Scalapino, R. L. Sugar,
N. E. Bickers, R. T.  Scalettar, Phys. Rev. B, {\bf 39}, 839 (1989).

\bibitem{moreo} A. Moreo and D. J. Scalapino, Phys. Rev. B {\bf 43}, 8211
(1991).

\bibitem{imada} Masatoshi Imada and Yasuhiro Hatsugai, J. Phys. Soc. Jpn.,
{\bf 58}, 3752 (1989).

\bibitem{zhang1} Shiwei Zhang, J. Carlson and J. E. Gubernatis, Phys. Rev.
Lett., {\bf 74}, 3652 (1995); Phys. Rev. B {\bf 55}, 7464 (1997);
J. Carlson, J. E. Gubernatis, G. Ortiz, and Shiwei Zhang, Phys. Rev. B 
{\bf 59}, 12788 (1999).

\bibitem{zhang2} Shiwei Zhang, J. Carlson and J. E. Gubernatis, Phys. Rev.
Lett., {\bf 78}, 4486 (1997).

\bibitem{scalettar1} R. T. Scalettar, Physica C, {\bf 162-164}, 313 (1989).

\bibitem{scalettar2} R. T. Scalettar, D. J. Scalapino, R. L. Sugar,
S. R. White, Phys. Rev. B {\bf 44}, 770 (1991).

\bibitem{dopf1} G. Dopf, A. Muramatsu, and W. Hanke, Phys. Rev. B
{\bf 41}, 9264 (1990).

\bibitem{dopf2} G. Dopf, A. Muramatsu, and W. Hanke, Phys. Rev. Lett.
{\bf 68}, 353 (1992).

\bibitem{dopf3} G. Dopf, J. Wagner, P. Dieterich, A. Muramatsu, and
W. Hanke, Helv. Phys. Acta {\bf 65}, 257 (1992).

\bibitem{dopf4} G. Dopf, A. Muramatsu, and W. Hanke,
Europhys. Lett. {\bf 17}, 559 (1992).

\bibitem{ogata} Masao Ogata and Hiroyuki Shiba, J. Phys. Soc. Jpn.,
{\bf 57}, 3074 (1988).

\bibitem{hirsch} J. E. Hirsch, S. Tang, E. Loh Jr., D. J. Scalapino,
Phys. Rev.  Lett. {\bf 60}, 1668 (1988); Phys. Rev. B {\bf 39}, 243
(1989).

\bibitem{stephan} W. H. Stephan, W. v. d. Linden and P. Horsch,
Phys. Rev. B {\bf 39}, 2924 (1989).

\bibitem{frick} M. Frick, P. C. Pattnaik, I. Morgenstern, D. M. Newns,
and W. v. d. Linden Phys. Rev. B {\bf 42}, 2665 (1990).

\bibitem{kuroki} Kazuhiko Kuroki and Hideo Aoki, Phys. Rev. Lett.
{\bf 76}, 4400 (1996).

\bibitem{comments} See also, H. Endres, W. Hanke, H. G. Evertz, 
and F. F. Assaad, Phys. Rev. Lett. {\bf 78}, 160 (1997); Kazuhiko
Kuroki and Hideo Aoki, Phys. Rev. Lett. {\bf 78}, 161 (1997).

\bibitem{guerr} M. Guerrero, J. E. Gubernatis, and Shiwei Zhang,
Phys. Rev. B {\bf 57}, 11980 (1998).

\bibitem{emery} V. J. Emery, Phys. Rev. Lett. {\bf 58},2794 (1987).

\bibitem{hybertsen} M. S. Hybertsen, M. Schuluter and N. E. Christensen,
Phys. Rev. B {\bf 39}, 9028 (1989).

\bibitem{eskes} H. Eskes, G. A. Sawatzky, L. F. Feiner, Physica C {\bf 160},
424 (1989).

\bibitem{mahan} A. K. McMahan,  J. F. Annett and R. M. Martin,
Phys. Rev. B {\bf 42}, 6268 (1990).

\bibitem{zhang} F. C. Zhang and T. M. Rice, Phys. Rev. B {\bf 37},
3795 (1988).

\bibitem{martin} Richard L. Martin, Phys. Rev. B {\bf 53}, 15501 (1996).

\bibitem{zaanen} J. Zaanen and A. M. Ole\'{s}, Phys. Rev. B {\bf 37}, 9423
(1988).

\bibitem{moreo1} Daniel Duffy and Adriana Moreo, Phys. Rev. B {\bf 52}, 15607
(1995).

\bibitem{moreo2} Charles Buhler and Adriana Moreo, Phys. Rev. B {\bf 59}, 9882
(1999).

\bibitem{mason} T. E. Mason, G. Aeppli, S. M. Haydeu, A. P. Hamiver, and
H. A. Mook, Phys. Rev. Lett. {\bf 71}, 919 (1993).

%\bibitem{furukawa} N. Furukawa and M. Imada, J. Phys. Soc. Jpn., {\bf 61},
%3331 (1992).
 
%\bibitem{husslein} T. Husslein, I. Morgenstern, D. M. Newns, P. C. Pattnaik,
% J. M. Singer, and H. G. Matuttis, Phys. Rev. B, {\bf 54}, 16179 (1996).
%
%\bibitem{Kuroki1} Kazuhiko Kuroki, Hideo Aoki, Takashi Hotta and Yasutami
%Takada, Phys. Rev. B, {\bf 55}, 2764 (1997).


\end{thebibliography}
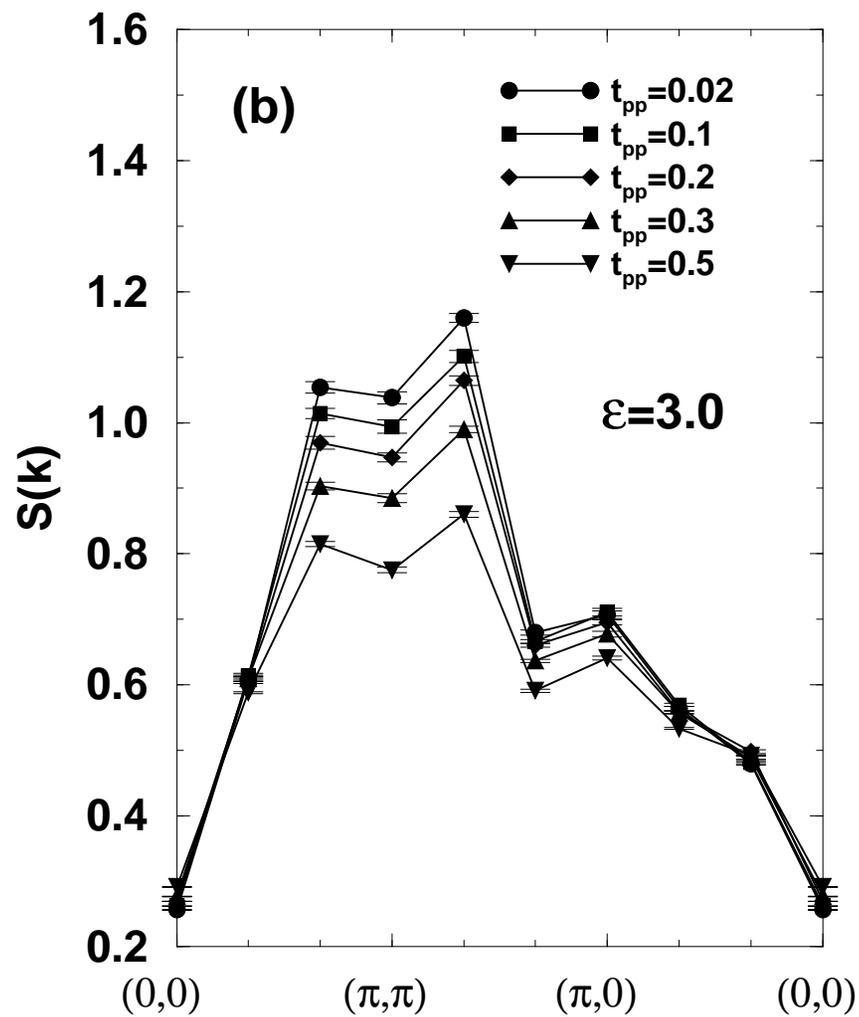
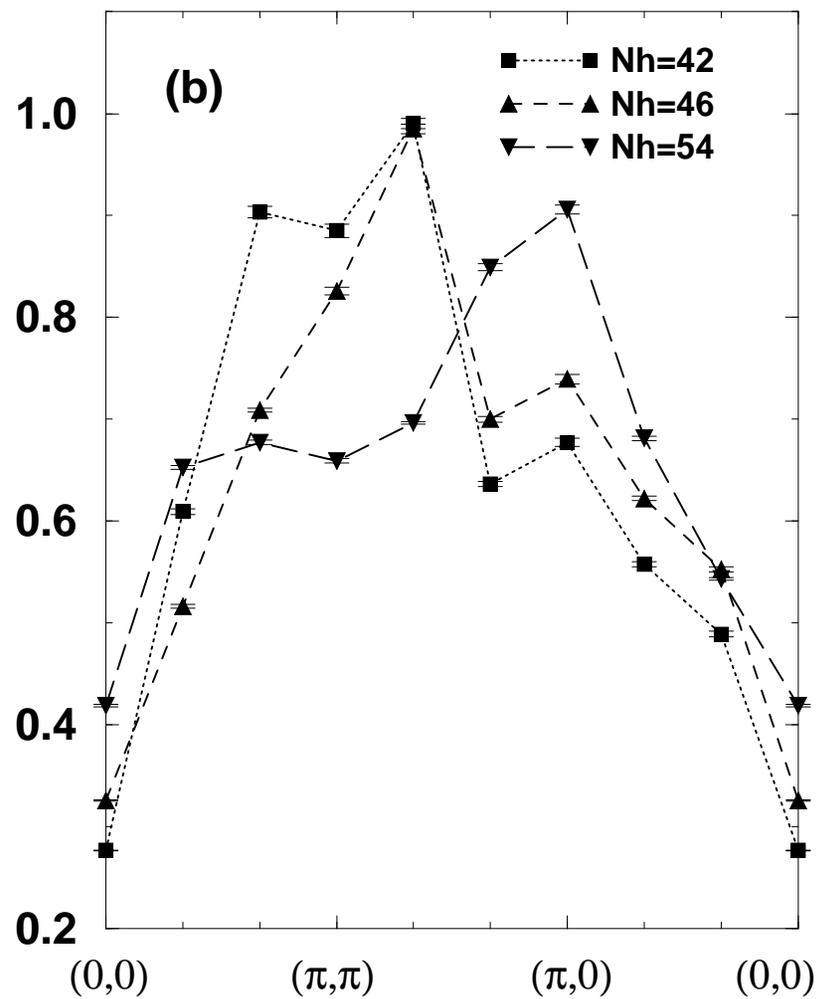
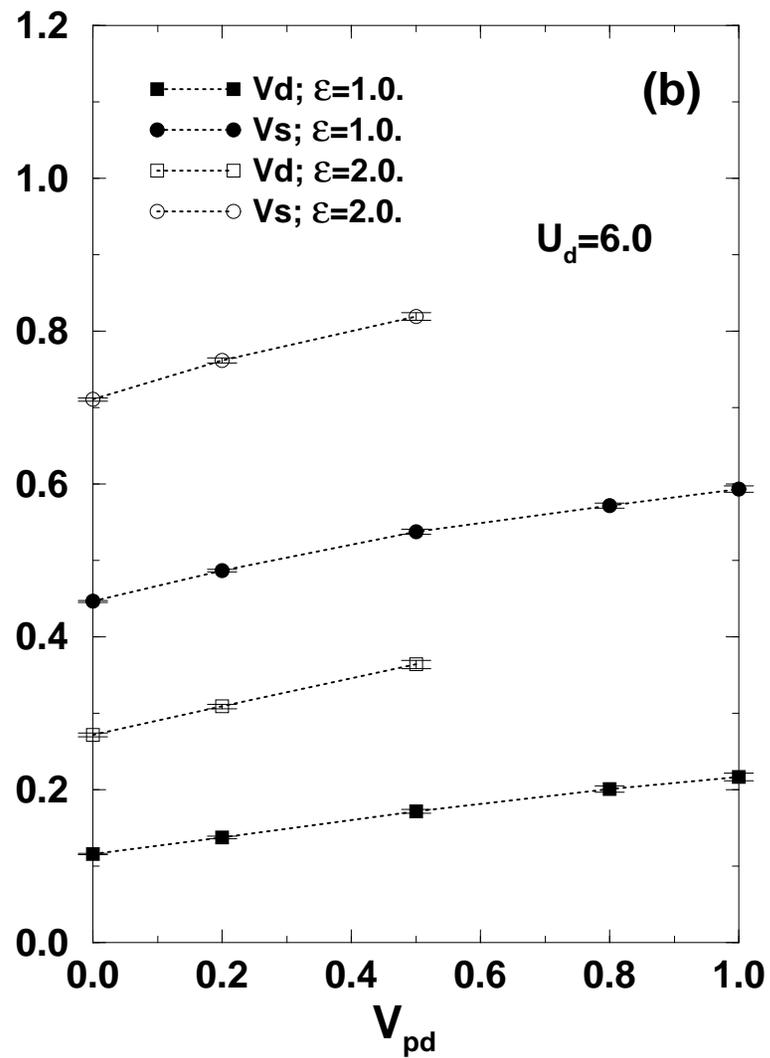
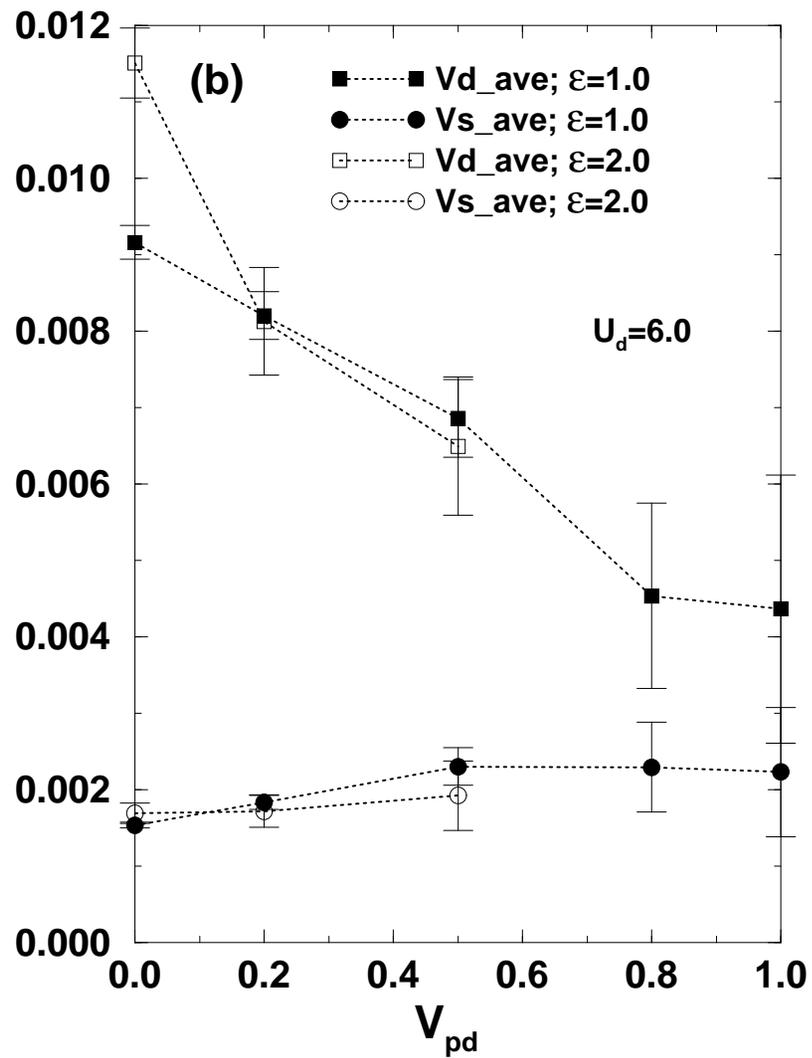
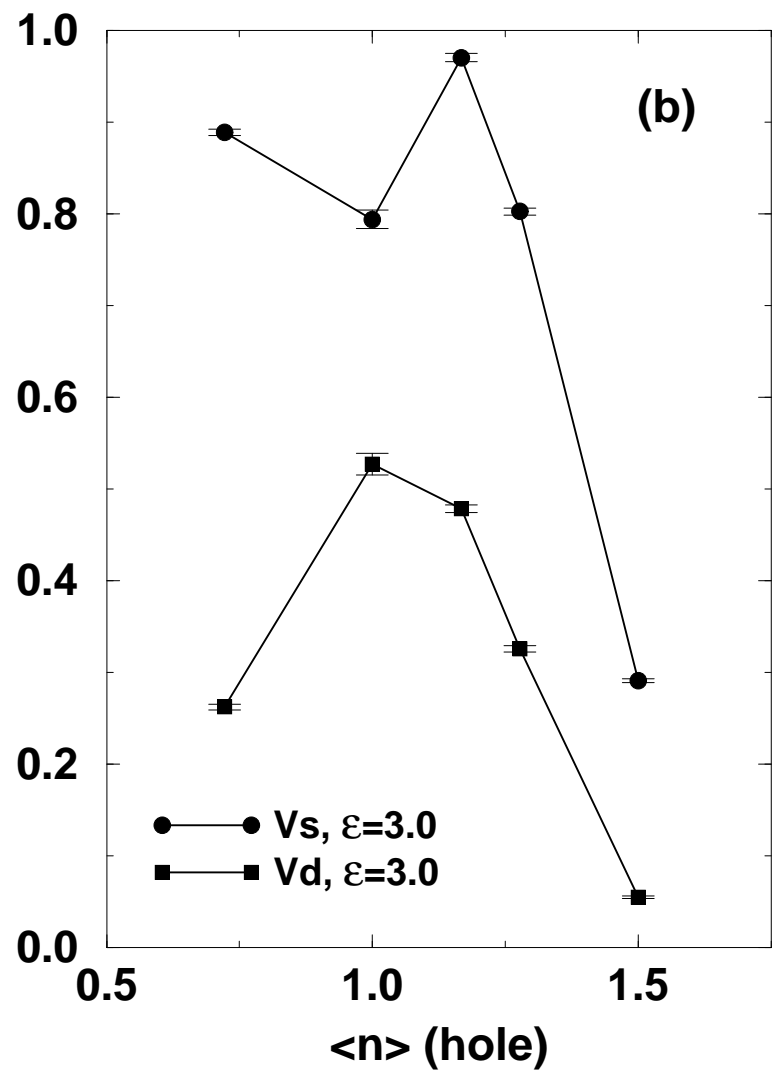
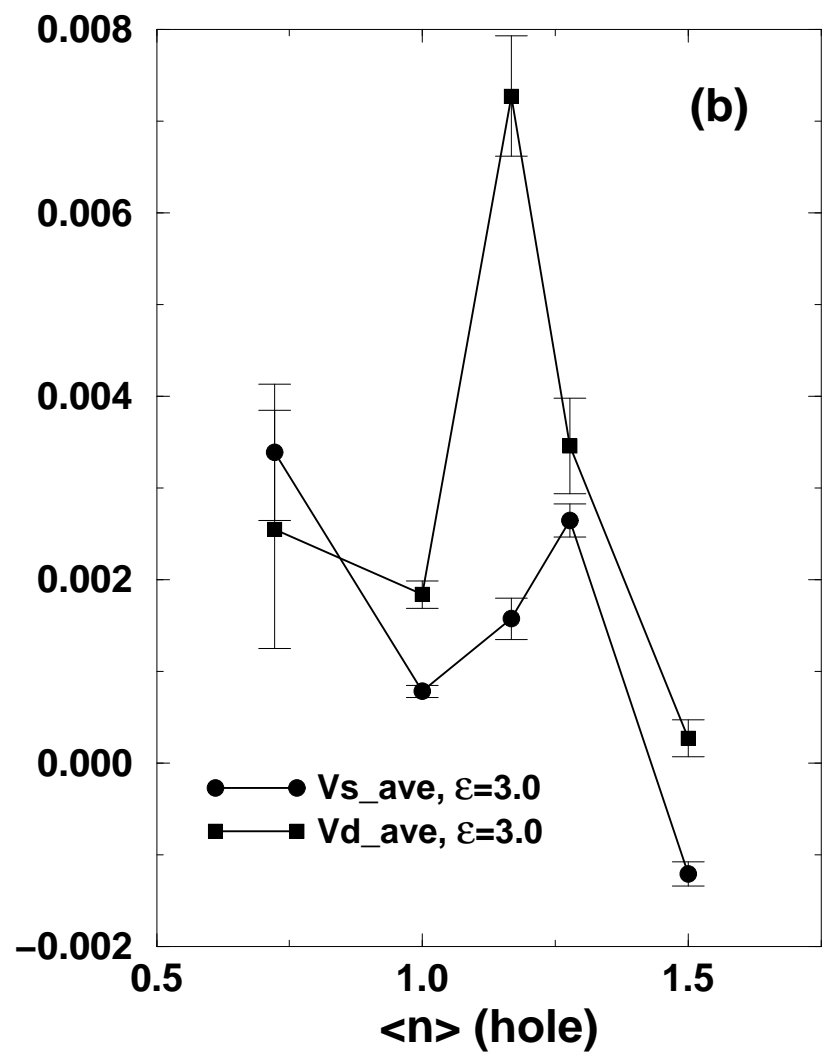
\end{document}